\documentclass{aa}  

\usepackage{graphicx}
\usepackage{caption}
\usepackage{subfigure}
\usepackage{float}
\usepackage{txfonts}
\usepackage{indentfirst}
\usepackage{comment}
\usepackage{siunitx}
\usepackage{booktabs}
\usepackage{dcolumn}
\usepackage{lscape}
\usepackage[table]{xcolor}
\usepackage{hyperref}
\usepackage{soul}
\DeclareSIUnit\year{yr}

\begin{document}

   \title{Parameter resolution of near-Earth asteroids using LISA}

   \subtitle{}

   \author{Sara Marques
          \inst{1,2}
          \and
          Oliver Jennrich\inst{3}
          }

   \institute{Space Research and Planetary Sciences, Physics Institute, University of Bern, Gesellschaftsstrasse 6, 3012 Bern, Switzerland
\and
Center for Space and Habitability, University of Bern, Gesellschaftsstrasse 6, 3012 Bern, Switzerland\\\email{sara.marques@unibe.ch}
         \and
             European Space Agency, ESTEC, Keplerlaan 1, 2201 AZ Noordwijk, The Netherlands\\
             \email{oliver.jennrich@esa.int}
             }

   \date{Received 5 January 2026; accepted 14 April 2026 }
 
  \abstract
   {The detection and tracking of near-Earth asteroids (NEAs) is critical for planetary defense and for advancing our understanding of the small-body population in the Solar System. Traditional detection techniques, including optical and radar, are the cornerstone of NEA discoveries. However, these are limited by observational constraints such as daylight, weather, and line-of-sight geometry.}
   {We investigate whether the Laser Interferometer Space Antenna (LISA), a future space-borne gravitational wave detector, might detect and constrain the parameters of NEAs through their gravitational effect on the spacecraft test masses during a close approach at the minimum orbital intersection distance.}
   {We modeled the gravitational perturbations of the test masses on the LISA spacecraft exerted by a passing NEA. Based on this, we computed its signal-to-noise ratio (S/N) to determine detectability using the LISA requirements. Furthermore, we used the Fisher information formalism to construct the Fisher matrix and estimate the uncertainties on the state vector and mass of the asteroid given it is detected. These estimates provide a first-order assessment of the LISA capability to resolve NEA parameters under favorable encounter geometries.}
   {The Fisher information study showed that the fractional uncertainty on the asteroid mass scales as the inverse of the S/N. Consequently, a detection with an S/N $\geq$ 5 yields a mass determination with an uncertainty of $\sim20\%$ at most. In contrast, the state vectors exhibit considerably larger uncertainties as they depend significantly on the geometry of the close approach. Nevertheless, for very high S/N cases, this precision may be comparable to the uncertainties obtained by some of the current observational methods. The analysis of the correlation matrices confirms that each close encounter produces a specific signal and provides a practical means to assign independent error bars to the recovered state vector. Additionally, this analysis reveals the emergence of recurring patterns linked to similar flyby geometries.
   }
   {}


    \clearpage

   \maketitle
%

\section{Introduction}\label{section:introduction}

Near-Earth asteroids (NEAs) are objects following trajectories around the Sun that are similar to that of Earth. They are of significant scientific interest because in studying them, we touch on key areas of planetary science, planetary defence, and space exploration. Despite its proximity, this population remains poorly characterized, especially in terms of physical properties such as their mass. According to \citep{Carry2012}, we can estimate the mass with an uncertainty of up to 10\% for less than 35\% of these asteroids. 

We study whether the Laser Interferometer Space Antenna (LISA) mission can measure the parameters of an asteroid, assuming that an encounter has already been detected by other means.
This novel characterization method for NEAs is insensitive to the prevailing observational biases and enhances our ability to characterize the intrinsic properties of these objects, particularly their masses. Section \ref{section:context} outlines the relevant background. In Section \ref{section:fisher_matrix} we describe the parameter-estimation method. In Section \ref{section_application} we present its application to a sample of asteroids from the observed population. Finally, we summarize and conclude this study in Section \ref{section:conclusion}.

\section{Context}\label{section:context}
\subsection{Near-Earth objects}

Near‑Earth objects are asteroids, comets, and other debris in the neighborhood of the Earth whose perihelion distances are smaller than \SI{1.3}{\astronomicalunit}. We restricted our analysis to near‑Earth asteroids because their orbital periods are shorter than those of comets, and they are therefore more likely to be observed during the operational lifetime of the LISA mission.

The detection of NEAs is strongly affected by survey‑dependent observational selection effects. The observed sample therefore does not represent the intrinsic population \citep{Granvik2018}. These effects arise from the underlying properties of the bodies (e.g., size, light‑curve amplitude, rotation period, apparent rate of motion, color, and albedo) and from survey characteristics (e.g., limiting magnitude, cadence, and exposure time). The completeness of NEA surveys depends on the diameter of the NEAs. According to \citep{Grav2023}, only 38\% of the NEAs with diameters larger than 140 \unit{\si{\metre}} have been discovered, whereas 88\% of those larger than 1 \unit{\si{\kilo\metre}} have been detected. Moreover, discoveries are strongly biased toward brighter, low‑inclination and low-eccentricity objects.

Some NEAs remain relatively unaltered since the formation of the Solar System. This makes them valuable archives of the primordial environment and events. Some of these were even studied to glean insights into early Solar System conditions and into the delivery of organic molecules and water ice to Earth. This information is relevant for determining the origin of life \citep{Ryugu}. Because of their proximity, NEAs also pose an impact hazard for Earth. An early detection and accurate orbit determination are essential for assessing collision risk and enabling mitigation strategies such as the deflection of an asteroid. 

\subsection{LISA}

The Laser Interferometer Space Antenna mission is a space mission led by the European Space Agency (ESA) with NASA as partner to detect and observe gravitational waves \citep{2024arXiv240207571C}. The currently planned launch date is July 2035. The measurement principle of LISA is based on high-precision interferometry to detect picometer-level changes in the distance between free-falling test masses (the arm length of the interferometer) as are caused by gravitational waves passing through the detector. The test masses are housed in pairs in three spacecraft, forming an approximately equilateral triangle. The frequency band in which LISA is most sensitive and in which the science measurements are taken extends from \SI{0.1}{mHz} to \SI{1}{Hz}. However, measurements at lower frequencies can be achieved, but are not necessarily protected against operational limitations.

The sensitivity of the instrument to changes in arm length makes LISA also susceptible to more conventional gravitational effects, such as the gravitational pull of asteroids passing by one of the LISA spacecraft (S/C). This gravitational pull can either be seen as a noise source for LISA, obscuring or at least degrading the gravitational wave signal, but can also be considered as a possibility to detect and characterize these asteroids.

The possibility that LISA can detect asteroids through the gravitational effect on the test masses was discussed earlier by \citep{2006CQGra..23.4939V} and also \citep{2009CQGra..26h5003T}. The main idea is to measure the acceleration of the test mass caused by the flyby of an asteroid as it manifests itself in the local measurement of the LISA mission, that is, a phase shift or a frequency shift in the laser reflected off the test mass.

While we took inspiration on the possible asteroid parameters from existing NEAs, we assumed an optimal measurement environment in which a LISA S/C and the asteroid approach each other as close as possible while still following their orbital path. We also simplified the problem by assuming that the asteroids are represented by point masses.

We furthermore ignored the self-gravity effect of a moving test mass. While the changed position of a test mass will affect the gravitational balance of the housing S/C, the position of the other test mass in the S/C, and ultimately, the orbit of the housing S/C, this effect is much smaller than the direct gravitational effect of a passing asteroid, and it plays out on timescales much longer than we studied. The long-term signature on the orbits still is a potential method for the detection of asteroids as well. For simplicity and computational efficiency, we also assumed a constant and equal arm length of the LISA constellation.

\subsubsection{Time-delay interferometry}

The sensitivities required for LISA can only be achieved when the frequency noise of the lasers used for the interferometric measurements can be sufficiently well reduced at the source or suppressed in the data analysis. Unlike ground-based gravitational wave detectors, for which the length of the interferometer arms is defined by construction, the distance between the LISA S/C changes over many thousands of kilometers over the lifetime of the mission due to orbital mechanics and the effect of Solar System bodies \citep{2021JAnSc..68..402M}. This means that frequency fluctuations of the laser appear to be many orders of magnitude larger in the readout than the gravitational wave signals. A method for significantly suppressing the frequency noise in post-processing is therefore required.  
 
The post-processing technique of time-delay interferometry (TDI) is employed in the LISA mission to achieve the required suppression of laser frequency noise. This TDI technique builds linear combinations of suitably time-delayed measurements in LISA to synthesize an equal-arm interferometer that is insensitive to frequency noise (see, e.g., \citep{2021LRR....24....1T} and references therein for a full discussion of TDI). This changes the time signature of the signals, however, and to properly assess the effect of the gravitational pull of asteroids, TDI post-processing must be applied to the signals. We used the \texttt{pyTDI} package, which implements TDI in Python \citep{pyTDI}.

The primary data streams produced by TDI, the so-called $X$, $Y$, and $Z$ channels, can be interpreted as the output of three equal-arm Michelson interferometers, with the beam splitter situated at each respective spacecraft. As each pair of these synthetic Michelson interferometers shares an arm, the noise correlation of the three channels is strong. However, three linear combinations $A$, $E$, and $T$ can be formed to minimize these correlations. Out of those three, only the $T$ channel has a simple interpretation as a Sagnac interferometer, which makes the $T$ channel less sensitive to low-frequency signals and suppresses the sensitivity to low-frequency signals significantly.
In Fig. \ref{fig:tdi_toutatis}, the signal in the $T$ channel is therefore displayed as amplified by a factor of 100 to render it visible.

\subsubsection{Signal of an asteroid}

\begin{figure}[htp!]
    \centering
    \includegraphics[width=1\linewidth]{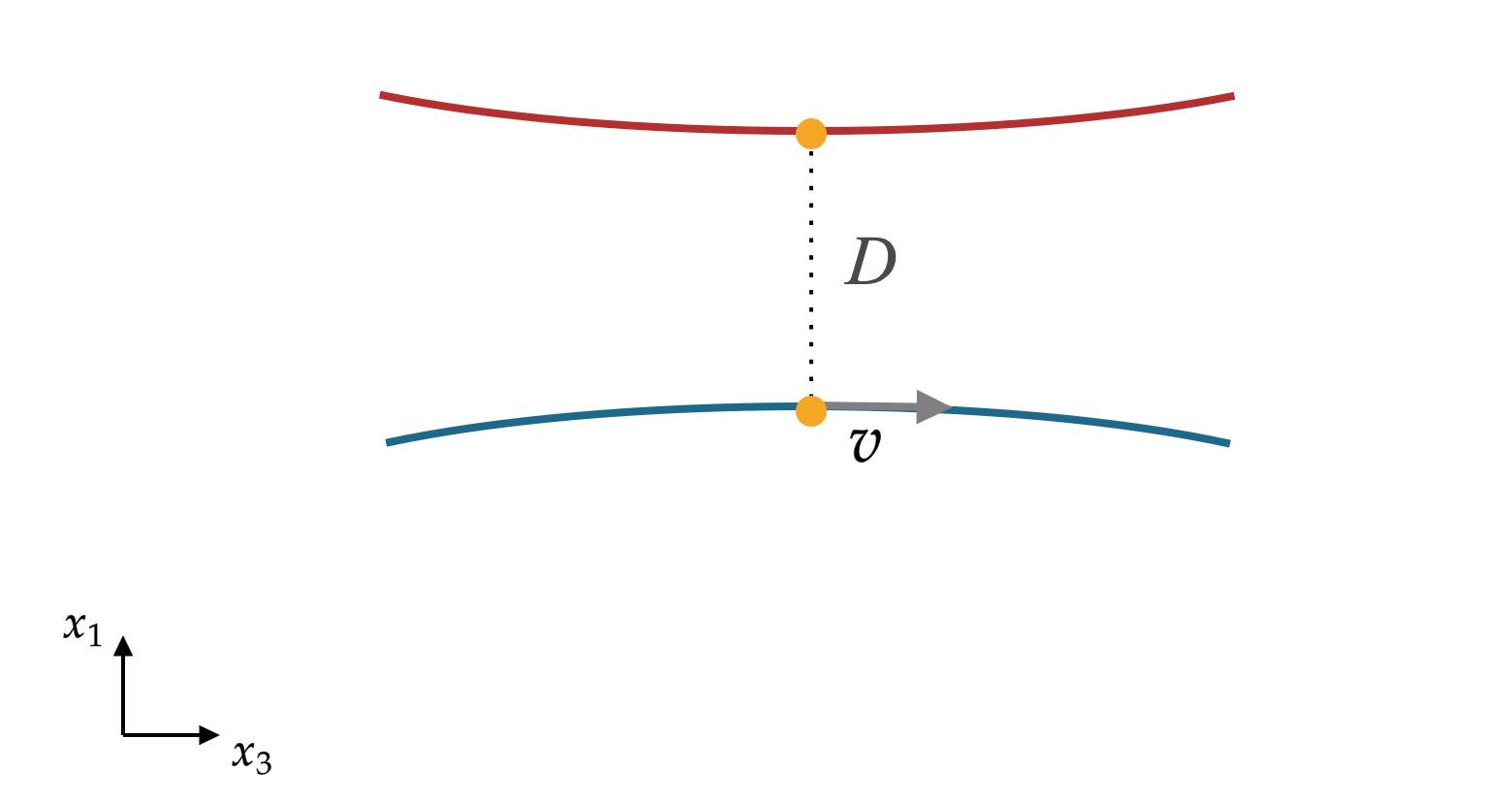}
    \caption{Geometry of a close approach in the reference frame ($x_1$, $x_3$), where a gravitational acceleration effect is induced on the test masses, as described in Eq.\ref{VinetEquation}. The blue line corresponds to the asteroid orbit, and the red line shows the LISA spacecraft. $D$ corresponds to the magnitude of the shortest distance between the two orbits, and $v$ presents the magnitude of the velocity of the asteroid when passing at the point of the shortest distance in its orbit.}
    \label{fig:moid2d}
\end{figure}

The gravitational acceleration of an asteroid on a test mass can be modeled as
\begin{equation}\label{VinetEquation}
\vec a(\tau) = \frac{Gm}{D^2}\frac{1}{\left(1+\tau^2\right)^{3/2}}\begin{pmatrix}1\\0\\\tau\end{pmatrix},
\end{equation}
where the coordinate system is chosen such that the third coordinate $x_3$ is given by the direction of the trajectory of the asteroid with respect to the test mass, and the first $x_1$ points in the direction of the shortest distance ($D$) between the spacecraft and the asteroid (see Fig. \ref{fig:moid2d}). The parameter $m$ is the mass of the asteroid, and $\tau = vt/D$ is the time $t$ scaled with the characteristic time $D/v$, that is, the shortest distance between the passing asteroid and the S/C, divided by the relative  velocity of the asteroid with respect to the S/C. The second coordinate $x_2$ is then chosen such that a right-handed coordinate system is formed.

\begin{figure}[htp!]
    \centering
    \includegraphics[width=1\linewidth]{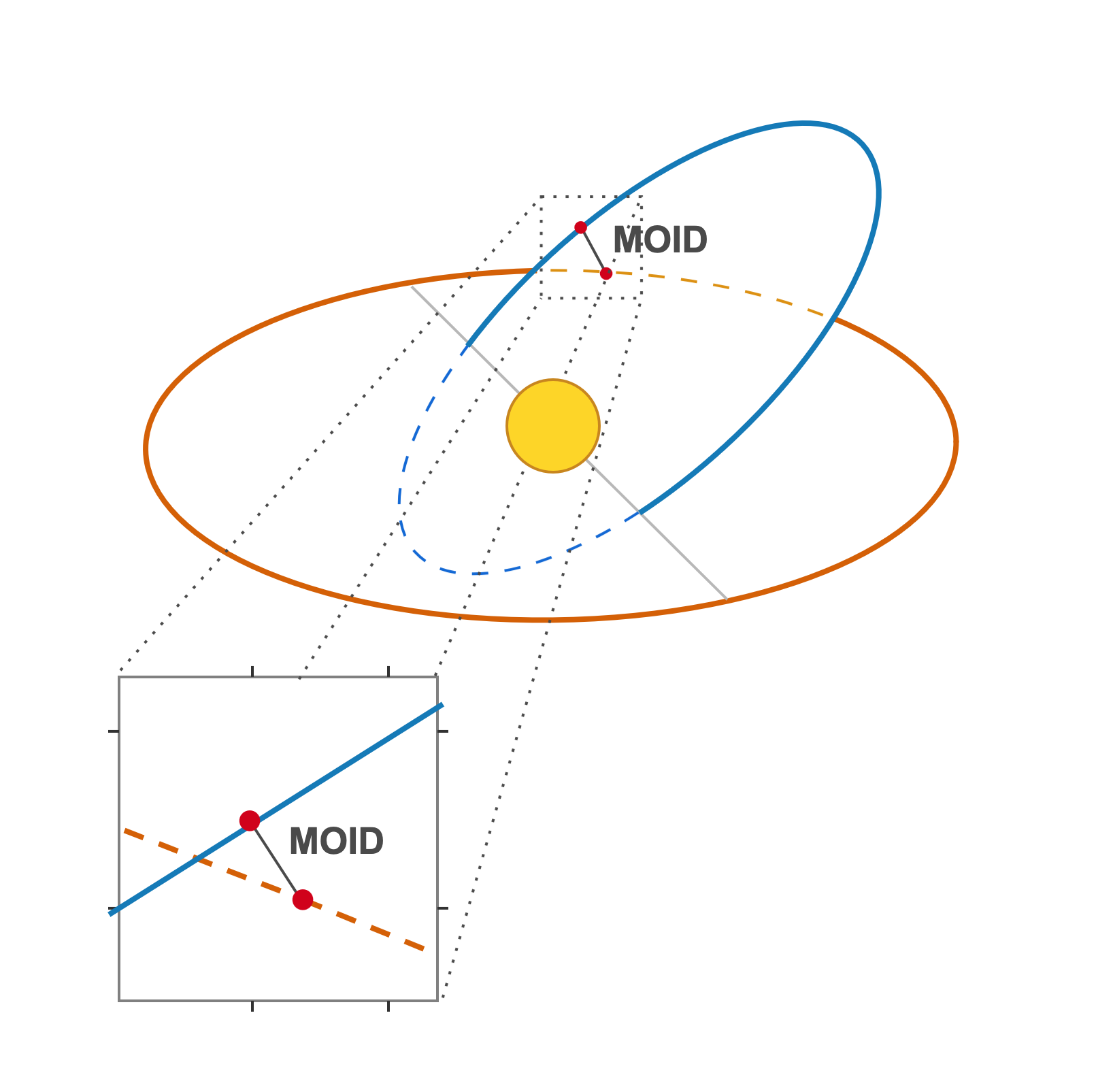}
    \caption{Geometry of the MOID concept for two conic sections in 3D space. The red dots correspond to the position of the two bodies in their respective orbit at which they would be at the smallest possible distance to each other.}
    \label{fig:moid_illustration}
\end{figure}

To maximize the signal, we created an optimal measurement by assuming that the asteroid and LISA S/C are at the minimum orbital intersection distance (MOID), corresponding to the minimum possible distance of the magnitude $D$ between the two orbits (shown in Fig. \ref{fig:moid_illustration}).
We adapted the work of \citep{Wisniowski2013}, who used an iterative approach to compute $D$, as well as the two mean anomalies at which two bodies approach the geometrical MOID position.

\begin{figure}[htp!]
    \centering
    \includegraphics[width=0.75\linewidth]{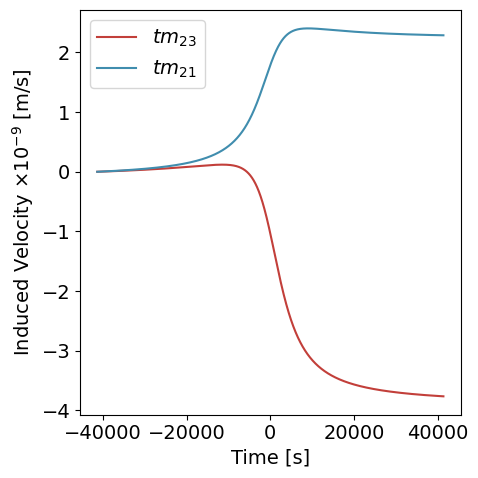}
    \caption{Velocity induced on the test masses during a simulated close approach between 4179 Toutatis and S/C 2 centered at $t = 0$ \unit{\si{\second}}. The test mass $tm_{ij}$ specifies the spacecraft $i$ to which it belongs and the spacecraft $j$ that is linked to it by the laser beam.
    For the other spacecraft, the velocity induced by the close approach is negligible and is thus not shown.}
    \label{fig:vel_ind_toutatis}
\end{figure}

Knowing the MOID, we simulated a close approach between the asteroid and one of the LISA spacecraft, and we collected the induced velocity on the test masses inside that spacecraft. An example of the resulting induced velocity for asteroid 4179 Toutatis is shown in Fig. \ref{fig:vel_ind_toutatis}.

The signal was then fed into the instrument simulator \texttt{LISA Instrument} (\citep{lisainstrument}, based on \citep{PhysRevD.107.083019}), which allowed us to imprint macroscopic movements onto the test masses in LISA. The movements were processed using the TDI technique, which generates a signal in the three TDI channels, $A$, $E$, and $T$. The measurements thus created were chosen to be in terms of fractional frequency changes, proportional to the velocity of the test masses. This obviated the introduction of an additional integration constant beyond the initial velocity (i.e., the initial position of the test mass in Eq. \ref{VinetEquation}) and simplified the analysis.
Furthermore, the resulting signal had the character of the derivative of the induced velocity on the test masses, which is due to the differencing scheme employed by TDI. At lowest order, the signal therefore had the appearance of the acceleration of the test mass. 

\begin{figure}[htp!]
    \centering
    \includegraphics[width=0.8\linewidth]{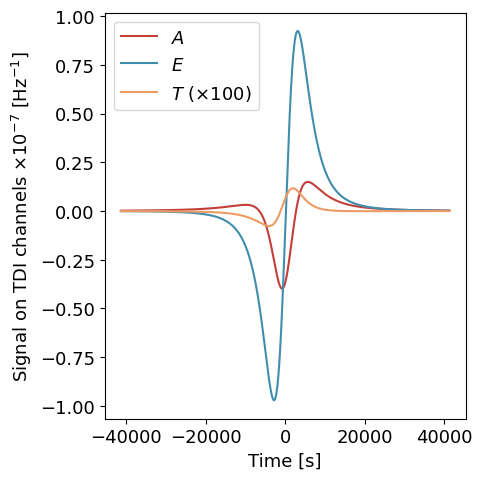}
    \caption{Signal on the TDI channels during a simulated close approach between 4179 Toutatis and S/C 2 centered at $t = 0$ \unit{\si{\second}}. Channel $T$ was enhanced by a factor of 100 for clarity.}
    \label{fig:tdi_toutatis}
\end{figure}

We used a sampling frequency of $f_s = \SI{1}{\hertz}$ for our test cases, which is commensurate with the upper band limit of LISA. An example is shown in Fig. \ref{fig:tdi_toutatis}. The signal was mostly concentrated in the first two channels ($A$ and $E$), while the signal in the $T$ channel was suppressed by a factor of about 100. As mentioned above, this is by construction, as the $T$ channel represents the fully symmetric signal in the LISA constellation, which is in the limit of very low frequencies that are insensitive to changes in the test-mass velocities (or positions). We therefore discarded the $T$ channel in our analysis.

To generate the signal, the mass of the asteroid was calculated under the assumption that it is a sphere of uniform density $\rho$ with a diameter $d$,

\begin{equation}\label{equation:mass}
    m = \frac{1}{6}\pi\rho d^{3}.
\end{equation}
We assumed an average density of $\rho=2200$ \unit{\si{\kilo\gram\per\metre^3}} to compute the asteroid mass based on \citep{Carry2012}, so that an asteroid with a diameter of 100 \unit{\si{\metre}} has a mass of about $1.15\times10^9$\,kg.

When the diameter was not provided by observations, we estimated it from the observed reflected light, assuming the asteroid to be a Lambertian (isotropic) scatterer in a shape of the disk in \unit{\si{\kilo\metre}} \citep{19940005152},

\begin{equation}
    d = \frac{1329}{\sqrt{p_\upsilon}}\,10^{-0.2H},
\end{equation}
where $p_\upsilon$ is the albedo, and $H$ is the absolute magnitude. When no albedo was measured for a specific asteroid, we used the intermediate albedo, $p_\upsilon = 0.14$ \citep{Pravec2012}, resulting in a resulting diameter of about \SI{500}{m} for an object with $H=15$.

To assess the detectability of a signal, we used the signal-to-noise ratio (S/N) definition from \citep{LISA-LCST-SGS-TN-001},

\begin{equation}
    \text{(S/N)}^2 = 4\int_{0}^{+\infty}\frac{|h(f)|^2}{S_n(f)} df,
\end{equation}
where $h(f)$ is the Fourier transform of the signal, and $S_n(f)$ is the power spectral density (PSD) of noise. We calculated the final S/N by adding the S/N for each individual TDI channel. We assumed  an $\text{S/N} \geq 5$ as a minimum detectability benchmark.

\subsection{LISA noise}

The expected LISA performance is defined in the LISA science requirements, and the relevant noise contributions were taken from \citep{LISA-LCST-SGS-TN-001}, 

\begin{equation}
    S_n(f) = \frac{1}{2}\frac{20}{3}\frac{(2\pi f)^2}{c^2}\left(\frac{S_I(f)}{(2\pi f)^4}+S_{II}(f)\right)\,R(f),
\end{equation}
for 0.1 \unit{\si{\milli\hertz}} $\leq f \leq$ 1 \unit{\si{\hertz}}, where $f$ is in \unit{\si{\hertz}}, and

\begin{align*}
       S_\mathrm{I}(f) &= 5.76\times 10^{-48}\left(1+\left(\frac{0.4 \,\unit{\si{\milli\hertz}}}{f}\right)^2\right)\\
       S_\mathrm{II}(f) &= 3.6\times 10^{-41}\\
       R(f) &= 1+\left(\frac{f}{25\,\unit{\si{\milli\hertz}}}\right)^2\\.
\end{align*}

 While the full noise model of LISA contains many individual contributions, the low-frequency noise ($S_I(f)$) is dominated by the acceleration noise of the test masses, whereas the high-frequency noise ($S_{II}$) is dominated by detection noise, mostly photon shot-noise. The instrument response function $R(f)$ approximately describes the effect of the finite arm length of LISA and the reduced sensitivity at high frequencies due to cancellation effects of the gravitational waves in the detector. The latter plays no role in the detection of the gravitational effects caused by asteroids as it only applies to signals accumulated by the light as it propagates between the spacecraft. It is only given for completeness.
The actual performance of LISA is likely to be slightly better as the performance requirements for the mission are about \SI{10}{\%} more stringent than the science requirements. The results presented here can therefore be considered the worst case.

\subsection{Simulation duration}
\label{sec:simulation_duration}

The duration of the simulation is a very important quantity to be defined and was taken to be a compromise between being able to sufficiently capture the essence of the signal while at the same time avoiding simulations that were too expensive computationally. The duration of the simulation was unique for each asteroid and depended on several parameters related to the close approach, such as the characteristic timescale $\tau$, the 3D geometry of the close approach, and the mass of the asteroid. 

In order to estimate the appropriate simulation time for each asteroid, we calculated the absolute induced acceleration on the spacecraft test mass for a long duration ($>$\SI{100000}{\second}), and we determined the time at which the absolute acceleration exceeded 2\% of the total acceleration range, that is,  the difference between maximum and minimum. The threshold of 2\% was chosen following a study of the effect of the threshold on the final calculated S/N for several asteroids.


\section{Fisher information}\label{section:fisher_matrix}

To be able to assess the parameter resolution, we adopted the Fisher information framework, which is commonly used by the gravitational wave community \citep{2008PhRvD.77h2008Q}, in the context of a maximum likelihood parameter estimation method. For a space of $\theta$ parameters to probe, the Fisher matrix elements $C^{-1}_{ij}$ can be calculated as

\begin{align}
    C^{-1}_{ij} &= \int_{-\infty}^{+\infty}\frac{\partial_{\theta_i}h(f)\,\partial_{\theta_j}h(f)^{*}+\partial_{\theta_i}h(f)^*\,\partial_{\theta_j}h(f)}{S_n(f)} \mathrm{d}f\\
    &= 4\int_0^\infty \frac{ \Re\left(\partial_{\theta_i}h(f)\,\partial_{\theta_j}h(f)^{*} \right)}{S_n(f)}\,\mathrm{d}f,
\end{align}
where $\partial_{\theta_i}h(f)$ is the partial derivative of the Fourier transform of the signal for the parameter $\theta_i$. 

The covariance matrix $C$ corresponds to the inverse of the Fisher matrix, where the diagonals are the variances of the signal for the parameters $\theta$. The correlation matrix $X$ can be found by applying a scaling matrix for normalization purposes on the covariance matrix. 

In our scenario, a close encounter between a NEA and a S/C can be seen as performing an astrometric measurement of its orbit, with the advantage that one of its physical properties is probed simultaneously. The parameters $\vec\theta_{\text{SV}}$ therefore correspond to its state vector at the MOID and the mass, 

\begin{equation}
    \mathbf{\vec \theta = \begin{pmatrix}
    x & y & z & v_x & v_y & v_z & m
                \end{pmatrix}^T}
\end{equation}
where $x$, $y$, and $z$ are the coordinates of the asteroid at the MOID in the barycentric celestial reference system, and $v_x$, $v_y$, and $v_z$ are its velocity.

In order to simplify the computations, we first computed the Fisher matrix and covariance matrix for the orbital elements $\vec\theta_\text{OE} = (a\quad e\quad i\quad \Omega\quad \omega\quad m)^T$, and we then sampled in a multivariate normal space defined by the MOID orbital parameters and the calculated covariances. We converted the new orbital elements into $\vec\theta_\text{SV} = (x\,\,\, y\,\,\, z\,\,\, v_x\,\,\, v_y\,\,\, v_z) $, which were used to deduce a new covariance matrix, standard deviations, and a correlation matrix for the state vector variables.
In our approach, the exact time of an encounter does not need to be identified to determine the uncertainties in $\theta_\text{OE}$. However, time does play a role in the conversion into the state vector, and determining the time $t_0$ of the encounter (and its statistics)  would require a proper detection pipeline. To avoid the additional complexity, we chose to estimate the uncertainty on the detection time by choosing $\sigma_t = T_\text{sig}/\text{(S/N)}$, where $T_\text{sig}$ is the duration of the signal as defined in Section \ref{sec:simulation_duration}. As most of the signal power is, by construction, contained in a shorter time interval, this estimation overestimates the timing uncertainty, and hence, the uncertainties on $\theta_\text{SV}$.

Furthermore, we added the contribution of each TDI channel to generate a global Fisher matrix. This allowed us to resolve the parameters better because summing the information of the different channels contributes to a lower covariance overall.


\section{Application to the observed population}\label{section_application}

We applied your method to the sample of the observed population of NEAs. To do this, we used the \texttt{LISABelt} package that we created, which implements a close approach between an asteroid and the LISA spacecraft in Python, including all the analysis methods we used  \citep{lisabelt}. The sample consisted of 20 asteroids\footnote{Their orbital elements and physical properties are listed in Table \ref{table:orbital_elements_NEA} in the appendix.} with an S/N above the minimum threshold for study collected from the database provided by the Near-Earth Objects Coordination Centre (NEOCC) from ESA. These were selected to provide balanced coverage spanning from a potential maximum LISA-detectable S/N (asteroid 4179 Toutatis) to the minimum threshold (S/N = 5). The results are highly sensitive to the assumed orbital parameters of the asteroid, meaning that small perturbations in these values can lead to significant changes in the outcomes.

In this section, we take the near-Earth asteroid 4179 Toutatis as an example to describe the process with which we obtained the parameter resolution using the Fisher information. We then show the results for the remaining sample. 4179 Toutatis was first seen in 1934, but was lost soon afterwards for decades, to be rediscovered again in 1989. It is a potentially hazardous asteroid (PHA) with the highest S/N and an excellent test case for computing the parameter resolution. 

To be able to calculate its Fisher information, we first computed the derivative of the signal for the parameters $\vec\theta_\text{OE}$. We selected S/C 2 for this close approach\footnote{We chose S/C 2 in this example because it has the lowest MOID with 4179 Toutatis (corresponding to MOID $=0.00041$ \unit{\si{\astronomicalunit}}), but the process is valid for any spacecraft.}.
The derivative was obtained using its definition, that is, by varying the parameters $\vec\theta_\text{OE}$ of the asteroid by a small delta $\vec\Delta$ (for this example, $\vec\Delta = 10^{-6}\,\vec\theta_{OE}$). We computed the difference between the signals and weighted it by the variation. Because of the unique geometry of the close approaches between the different asteroids and the LISA spacecraft, the effect of varying the different parameters on the signal differs from one asteroid to the next. From this, we determined the Fisher elements $C_{ij}^{-1}$ of 4179 Toutatis  for the $\vec\theta_{OE}$ parameters, 

{\footnotesize
\begin{equation}\label{FM_orbital_elements}
    C^{-1}_{OE} = 10^{11}\begin{pmatrix}
        0.09 & -0.57 & -85.42 & 0.30 & 0.04 \\
        -0.57 & 4.06 & 558.56 & -1.99 & -0.29 \\
        -85.42 & 558.56 & 8.70\times{10^4} & -307.26 & -42.80\\
        0.30 & -1.99 & -307.26 & 1.09 & 0.15 \\
        0.04 & -0.29 & -42.80 & 0.15 & 0.02
    \end{pmatrix}
\end{equation}
}

{\footnotesize
\begin{equation}\label{FM_mass}
    C^{-1}_{m} = 10^{-18}\times 1.08,
\end{equation}
}where the semimajor axis is given in \unit{\si{\astronomicalunit}}, the angles in \unit{\si{\radian}}, and the mass in \unit{\si{\kilo\gram}}.

We treated the mass separately from the other parameters because it is a scaling parameter, and it therefore does not change the shape of the signal. In particular,

\begin{align}
    \nonumber\partial_m h(f) &= \frac{1}{m} h(f)\\
    C_{mm} &= \frac{1}{m^2}\text{(S/N)}^2
\end{align}
and
\begin{align}
    \nonumber\partial_m h(f) \partial_j h(f) &= \frac{1}{m} h(f) \partial_j h(f)\\
    \nonumber&= \frac{1}{2m}\partial_j h(f)^2\\
    C_{mj} &=\frac{1}{2m} \partial_j \text{(S/N)}^2,
\end{align}
allowing us to directly calculate the relative matrix elements from the S/N. These $C_{mj}$ elements are lower by several orders of magnitude than the covariances between the orbital elements themselves. Therefore, we validate the hypothesis that the mass is not correlated with the orbital elements and can thus be treated separately without affecting the results.

We inverted the Fisher matrix $C^{-1}_\text{OE}$ to obtain the covariance matrix $C_{OE}$ of the orbital parameters $\vec\theta_\text{OE}$. This covariance matrix provided the covariances of multivariate normal distributions in a space defined by the $\vec\theta_\text{OE}$ parameters. The mean for these multivariate normals corresponds to the actual orbital elements of the asteroid. We therefore sampled the multivariate normal distributions to obtain a covariance matrix for the desired state vector parameters $\vec\theta_\text{SV}$. For 4719 Toutatis, the covariance matrix for the state vector $C_\text{SV}$ is given by

{\tiny
\begin{equation}
    C_{SV} = 10^{-8}\begin{pmatrix}
        0.59 & -0.47 & 0.09 & 23.83 & 18.69 & 6.47\\
        -0.47 & 0.43 & -0.07 & -19.70 & -16.83 & -4.71\\
        0.09 & -0.07 & 0.02 & 3.61 & 2.72 & 1.72\\
        23.83 & -19.70 & 3.61 & 979.20 & 784.20 & 258.58\\
        18.69 & -16.83 & 2.72 & 784.20 & 657.77 & 193.53\\
        6.47 & -4.71 & 1.72 & 258.58 & 193.53 & 124.31\\
    \end{pmatrix},
\end{equation}
}where the distances are provided in \unit{\si{\astronomicalunit}} and the velocities in \unit{\si{\kilo\meter\per\second}}.

The standard deviations for the asteroid parameters can be directly obtained from the covariance matrix $C_{SV}$ for the state vector, and they can be obtained by inverting the Fisher matrices $C^{-1}_{m}$ for the mass. We extracted the diagonals of the covariance matrices and computed their square root to obtain the respective standard deviations. The retrieved results for 4179 Toutatis are shown in Table \ref{table:standard_deviations_toutatis}.

\renewcommand{\arraystretch}{1.25}
\begin{table}[htp!]
{
\centering
\caption{Standard deviations obtained using the Fisher information method for asteroid 4179 Toutatis and comparison with observations.}
{
\begin{tabular}{@{}ccc@{}}
\toprule
  Standard Deviations & Observations & This Work  \\ 
\cmidrule(r){1-1}\cmidrule(lr){2-2}\cmidrule(l){3-3}
 $\sigma_x$ [\unit{\si{\astronomicalunit}}] & $4.13\times10^{-9}$ & $7.65\times10^{-5}$ \\ 
 $\sigma_y$ [\unit{\si{\astronomicalunit}}] & $9.25\times10^{-9}$ & $6.59\times10^{-5}$ \\
 $\sigma_z$ [\unit{\si{\astronomicalunit}}] & $3.05\times10^{-9}$ & $1.54\times10^{-5}$\\
 $\sigma_{v_x}$ [\unit{\si{\kilo\meter\per\second}}] & $2.27\times10^{-7}$ & $3.13\times10^{-3}$\\
 $\sigma_{v_y}$ [\unit{\si{\kilo\meter\per\second}}] & $3.27\times10^{-7}$ & $2.56\times10^{-3}$\\
 $\sigma_{v_z}$ [\unit{\si{\kilo\meter\per\second}}] & $8.96\times10^{-8}$ & $1.11\times10^{-3}$ \\
  $\sigma_m^{*}$ [\%] & -- & $4.55\times10^{-3}$ \\
\bottomrule
\end{tabular}}
\tablefoot{The standard deviations for the observations were collected from the \href{https://neo.ssa.esa.int}{ESA NEOCC website} at epoch 61200\, mjd. The mass standard deviation, $\sigma_m^*$, corresponds to the normalized version $\sigma_m^* = \sigma_m/m$ by the mass $m$ of asteroid 4179 Toutatis.}
\label{table:standard_deviations_toutatis}
}
\end{table}

While the uncertainty in the state vector is considerably larger than in current observations, we note that the observational uncertainty is a result of more than 7000 astrometric and radar observations since 1992, while LISA would only have measured it once. This translates into an angular precision\footnote{The angular precision (in arcseconds) was computed from the Earth-LISA spacecraft distance (54 million \unit{\si{\kilo\meter}} $\pm$ MOID distance) and the coordinate uncertainties of the state vector. The maximum precision was calculated using the minimum distance and largest coordinate uncertainty. The minimum precision employed the maximum distance and smallest coordinate uncertainty.} of $26.06\pm17.35$ \unit{\si{\arcsecond}}. Although this is somewhat higher than typical astrometric measurements \citep{VERES2017139}, it can provide the means for determining a preliminary orbit. Classically, this requires at least three well-spaced positional measurements. The LISA measurement provides sufficient information for this in one very short observation, which represents a qualitatively different capability that is complementary to traditional astrometry. The temporal uncertainty of 3.77 seconds relative to a signal duration of approximately 20,000 seconds indicates that if such an event were to occur, the epoch of closest approach could be determined with high precision, and we would be able to extract an accurate state vector. While 4179 Toutatis represents a high-S/N case and is unsuitable as a general benchmark, our results suggest that detections of previously unknown NEAs in this scenario might yield sufficient information for a preliminary orbit determination from a single measurement. This would allow a potential follow-up later. Furthermore, due to the high S/N and the inherent sensitivity of LISA to gravitational effects, a close approach with asteroid 4179 Toutatis would achieve an unprecedented mass determination with exceptional precision.

\begin{figure}[htp!]
    \centering
    \includegraphics[width=0.8\linewidth]{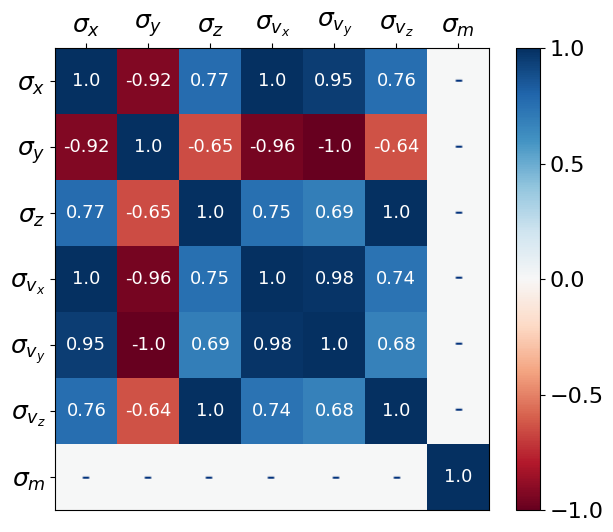}
    \caption{Correlation matrix of the uncertainties on the parameter $\theta$ for asteroid 4179 Toutatis. The covariances between the mass and other state vector variables are not assessed.}
    \label{fig:correlation_matrix_toutatis}
\end{figure}

The examination of its correlation matrix (Fig. \ref{fig:correlation_matrix_toutatis}) notably reveals stronger correlations of the x- and y-axis components than of the z-axis component. This pattern is consistent with the low orbital inclination of the asteroid, which constrains its motion predominantly to the ecliptic plane. Consequently, the x and y coordinate and velocity components are substantially coupled with each other, such that uncertainties in one component propagate to the others. However, these correlations are intrinsic to the specific close-approach geometry of 4179 Toutatis. Alternative close-encounter configurations may produce different interdependences among the state vector components.

In Table \ref{table:standard_deviations_all} we list the standard deviations for the other sample asteroids. As expected, the uncertainty on the parameter determination decreases with increasing S/N.
In particular, the fractional standard deviation of the mass $\sigma_m/m$ is proportional to the inverse of the S/N, that is, $\sigma_m^{*} \propto \text{(S/N)}^{-1}$. As the mass scales the signal amplitude (see Eq. \ref{VinetEquation}), meaning $m \partial_m \,h(f) = h(f)$, we have that
\begin{equation}
    C^{-1}_{mm} = \frac{1}{m^2}\text{(S/N)}^2 \Rightarrow \sigma_m^* = \frac{\sigma_m}{m} = \frac{1}{\text{(S/N)}}.
\end{equation}
It is therefore possible to directly estimate the standard deviation for the mass from the S/N. This is not always the case for the standard deviations of the state vector because they change according to the geometry of the flyby, and thus, the shape of the signal changes. Furthermore, for only very low S/N, we have uncertainties of about a few percent for the mass. This means that if an asteroid is detected by LISA (S/N $\geq$ 5), we will be able to determine its mass with good precision (20\% or lower), whereas a significantly higher S/N is required to determine the state vector with high fidelity. Furthermore, the uncertainty on the time parameter is typically about a few seconds to some tenths of a second (for an S/N > 100). In this regime, the signals can persist for several hours. The temporal precision therefore implies that when a close approach occurs and is detectable in the data, its timing can be determined with considerable accuracy. Only in a regime with a low signal-to-noise ratio (S/N $\leq$ 100), the uncertainty on the time of closest approach become non-negligible. This might limit our ability to accurately constrain the epoch of the event.

The analysis of the uncertainty on the state vector of asteroid 2015 DU180 revealed an important limitation of our study: a high eccentricity combined with a low S/N can yield unphysical parabolic or hyperbolic orbital solutions that lead to substantial velocity uncertainties. While this issue is inherent to high-eccentricity objects, it affects low-eccentricity orbits only little because a small MOID distance is necessary for detectability (due to the near-circularity), which ensures a high S/N. For low eccentricities, the close-approach geometry can therefore be determined well, and this prevents unrealistic (negative) eccentricity values.

In practice, LISA state vector measurements are expected to achieve typical accuracies on the order of hundreds of arcseconds, with worst-case scenarios reaching thousands of arcseconds. For previously observed asteroids, this precision would provide limited improvements of the orbit determination, with the primary scientific value residing in mass measurements. For asteroids detected by LISA for the first time, the positional uncertainty would be insufficient for a direct orbital solution, but it could delineate a somewhat constrained search region in the sky, facilitating targeted follow-up observations with ground- or space-based facilities. 

\begin{figure}[htp!]
    \centering
\begin{minipage}[c]{\textwidth}
    \centering
    \begin{minipage}[c]{0.19\textwidth}
        \centering
        \includegraphics[width=1.9cm]{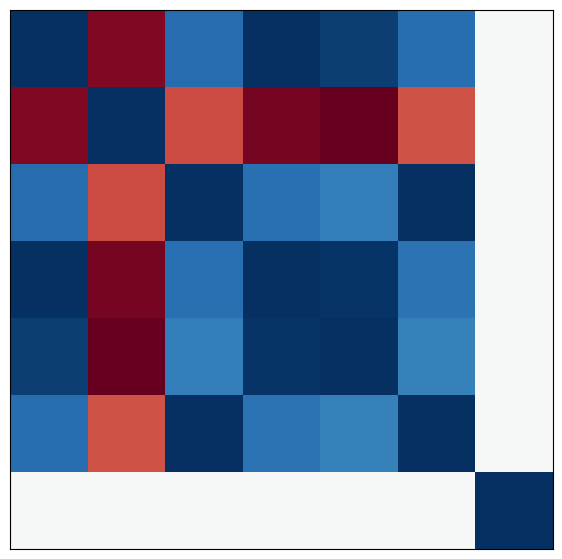}
        \captionsetup{labelformat=empty}
        \vspace{-0.2cm}
        \captionof{subfigure}{4179 Toutatis}\label{fig:4179_toutatis_corr}
    \end{minipage}
    \begin{minipage}[c]{0.19\textwidth}
        \centering
        \includegraphics[width=1.9cm]{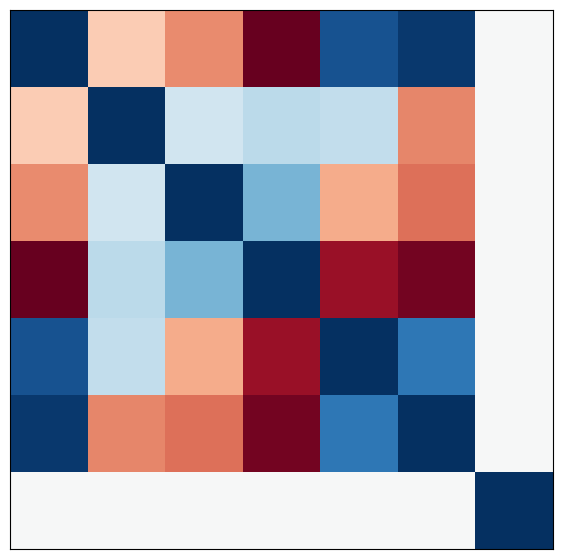}
        \captionsetup{labelformat=empty}
        \vspace{-0.2cm}
        \captionof{subfigure}{69230 Hermes}\label{fig:69230_Hermes_corr}
    \end{minipage}
    \begin{minipage}[c]{0.19\textwidth}
        \centering
        \includegraphics[width=1.9cm]{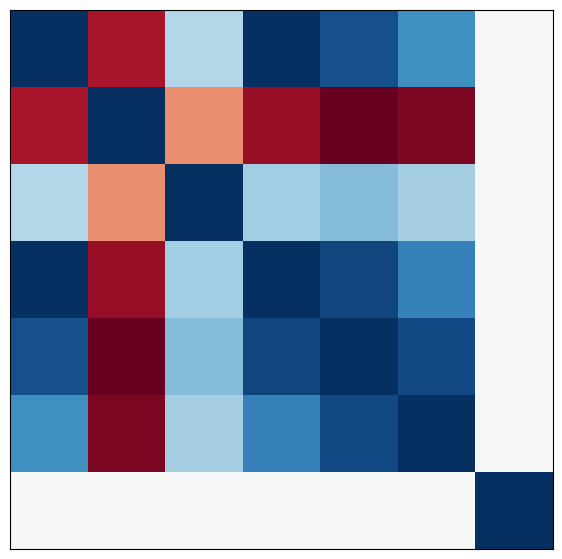}
        \captionsetup{labelformat=empty}
        \vspace{-0.2cm}
        \captionof{subfigure}{2002 PZ39}\label{fig:2002_PZ39_corr}
    \end{minipage}
    \begin{minipage}[c]{0.19\textwidth}
        \centering
        \includegraphics[width=1.9cm]{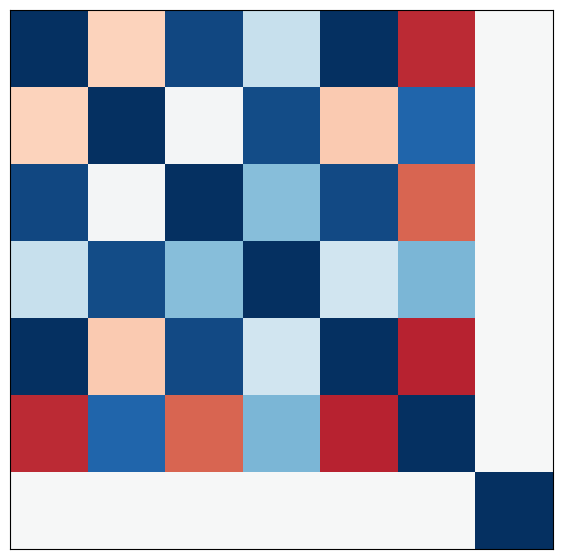}
        \captionsetup{labelformat=empty}
        \vspace{-0.2cm}
        \captionof{subfigure}{2013 FA8}\label{fig:2013_FA8_corr}
    \end{minipage}
    \begin{minipage}[c]{0.19\textwidth}
        \centering
        \includegraphics[width=1.9cm]{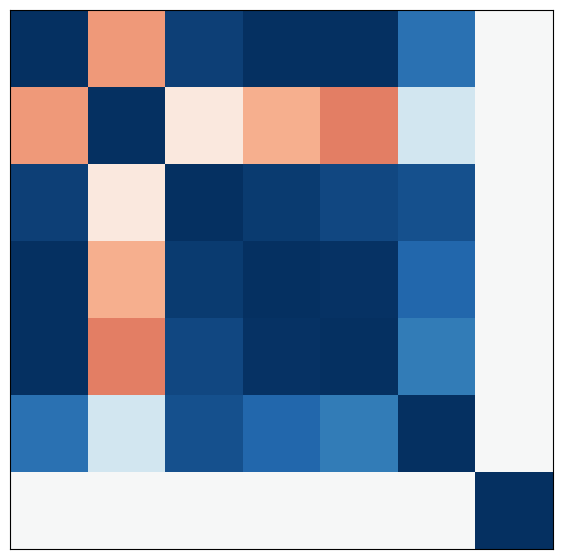}
        \captionsetup{labelformat=empty}
        \vspace{-0.2cm}
        \captionof{subfigure}{1997 BR}\label{fig:1997_BR_corr}
    \end{minipage}

    \vspace{0.15cm}
    \begin{minipage}[c]{0.19\textwidth}
        \centering
        \includegraphics[width=1.9cm]{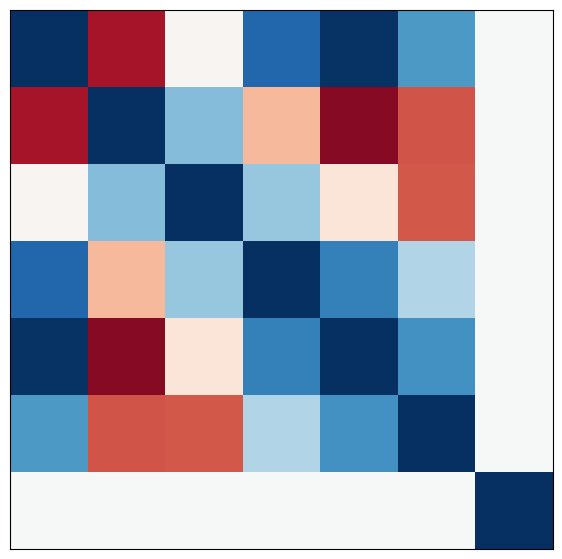}
        \captionsetup{labelformat=empty}
        \vspace{-0.2cm}
        \captionof{subfigure}{2007 UY1}\label{fig:2007_UY1_corr}
    \end{minipage}
    \begin{minipage}[c]{0.19\textwidth}
        \centering
        \includegraphics[width=1.9cm]{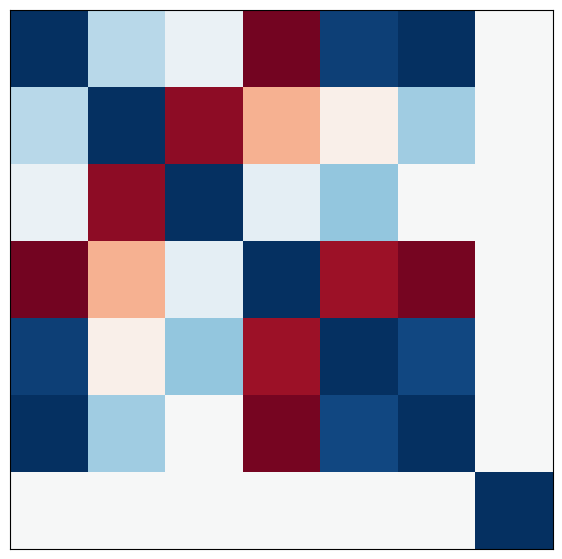}
        \captionsetup{labelformat=empty}
        \vspace{-0.2cm}
        \captionof{subfigure}{2003 QQ47}\label{fig:2003_QQ47_corr}
    \end{minipage}
    \begin{minipage}[c]{0.19\textwidth}
        \centering
        \includegraphics[width=1.9cm]{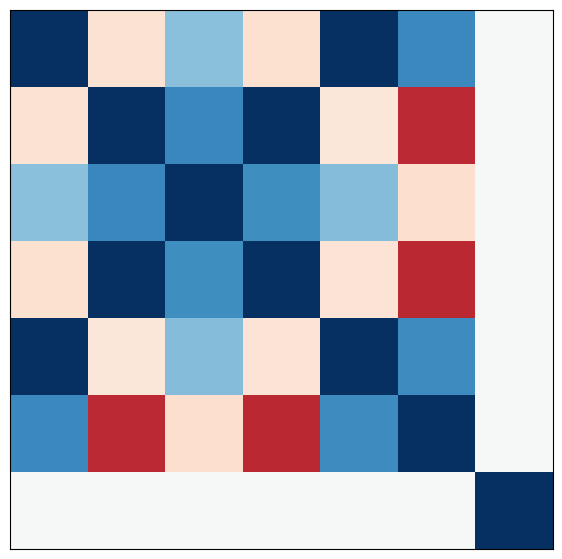}
        \captionsetup{labelformat=empty}
        \vspace{-0.2cm}
        \captionof{subfigure}{1999 MN}\label{fig:1999_MN_corr}
    \end{minipage}
    \begin{minipage}[c]{0.19\textwidth}
        \centering
        \includegraphics[width=1.9cm]{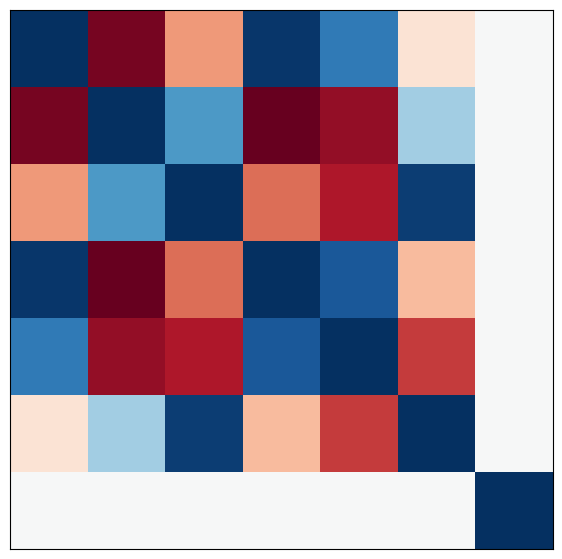}
        \captionsetup{labelformat=empty}
        \vspace{-0.2cm}
        \captionof{subfigure}{2003 YK118}\label{fig:2003_YK118_corr}
    \end{minipage}
    \begin{minipage}[c]{0.19\textwidth}
        \centering
        \includegraphics[width=1.9cm]{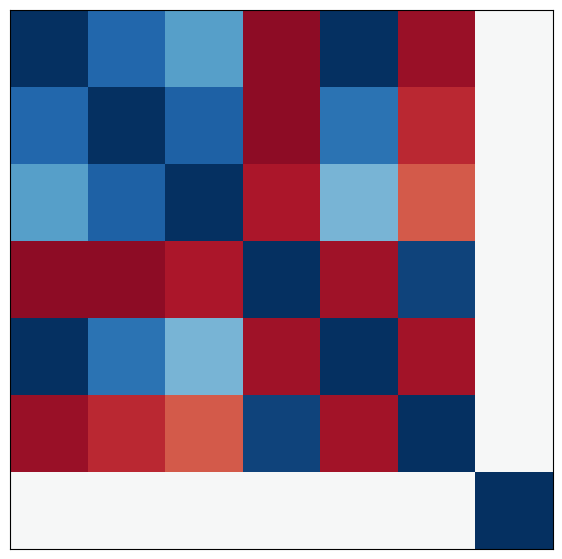}
        \captionsetup{labelformat=empty}
        \vspace{-0.2cm}
        \captionof{subfigure}{2011 BO24}\label{fig:2011_BO24_corr}
    \end{minipage}

    \vspace{0.15cm}
    \begin{minipage}[c]{0.19\textwidth}
        \centering
        \includegraphics[width=1.9cm]{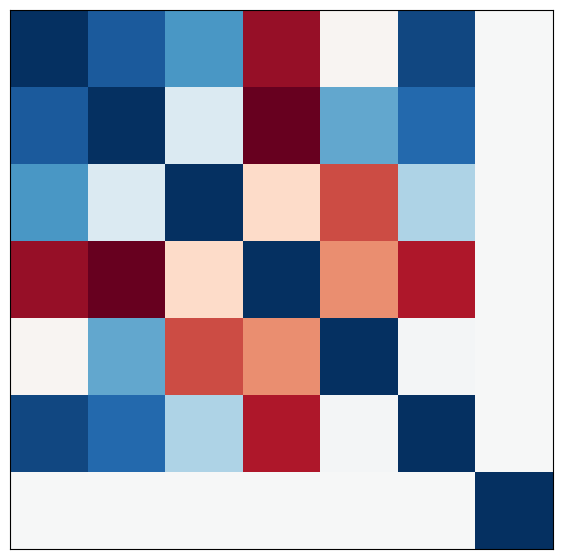}
        \captionsetup{labelformat=empty}
        \vspace{-0.2cm}
        \captionof{subfigure}{2003 AC23}\label{fig:2003_AC23_corr}
    \end{minipage}
    \begin{minipage}[c]{0.19\textwidth}
        \centering
        \includegraphics[width=1.9cm]{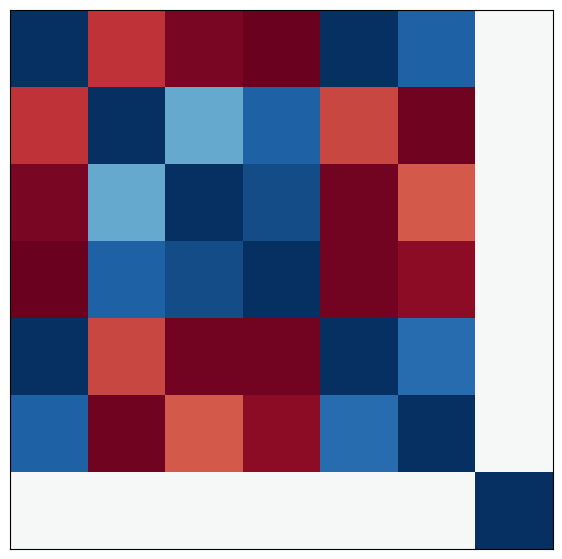}
        \captionsetup{labelformat=empty}
        \vspace{-0.2cm}
        \captionof{subfigure}{2020 FR1}\label{fig:2020_FR1_corr}
    \end{minipage}
    \begin{minipage}[c]{0.19\textwidth}
        \centering
        \includegraphics[width=1.9cm]{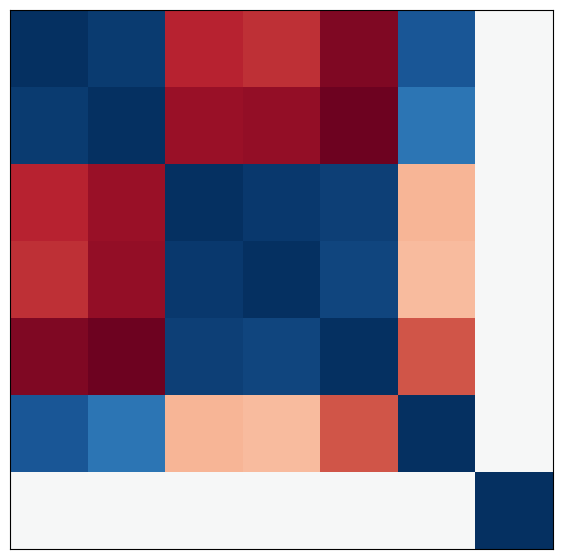}
        \captionsetup{labelformat=empty}
        \vspace{-0.2cm}
        \captionof{subfigure}{2018 DA1}\label{fig:2018_DA1_corr}
    \end{minipage}
    \begin{minipage}[c]{0.19\textwidth}
        \centering
        \includegraphics[width=1.9cm]{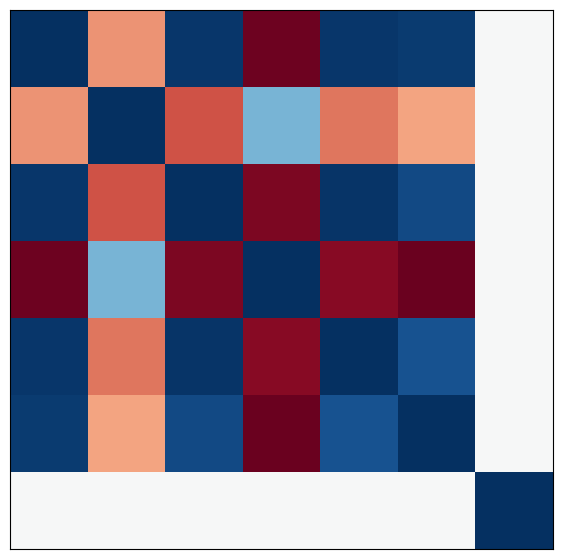}
        \captionsetup{labelformat=empty}
        \vspace{-0.2cm}
        \captionof{subfigure}{2016 CU193}\label{fig:2016_CU193_corr}
    \end{minipage}
    \begin{minipage}[c]{0.19\textwidth}
        \centering
        \includegraphics[width=1.9cm]{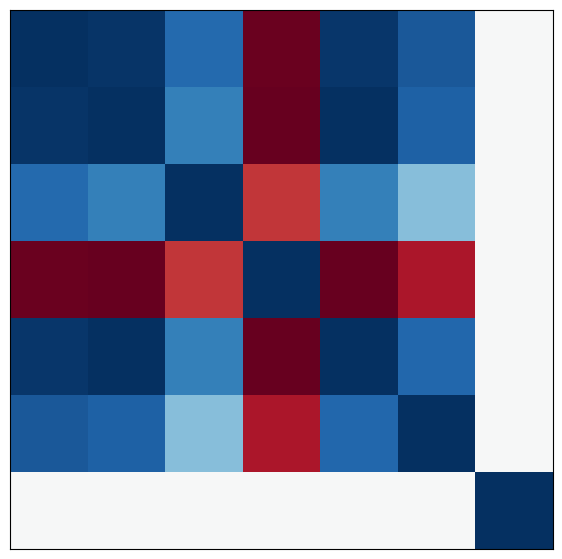}
        \captionsetup{labelformat=empty}
        \vspace{-0.2cm}
        \captionof{subfigure}{2014 UC192}\label{fig:2014_UC192_corr}
    \end{minipage}

    \vspace{0.15cm}
    \begin{minipage}[c]{0.19\textwidth}
        \centering
        \includegraphics[width=1.9cm]{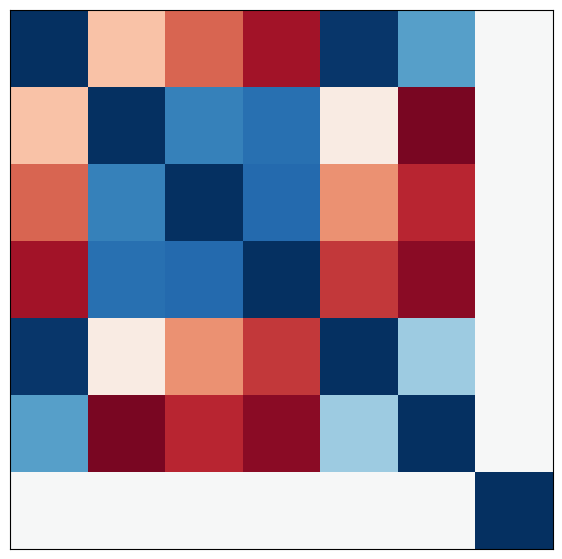}
        \captionsetup{labelformat=empty}
        \vspace{-0.2cm}
        \captionof{subfigure}{2005 TU50}\label{fig:2005_TU50_corr}
    \end{minipage}
    \begin{minipage}[c]{0.19\textwidth}
        \centering
        \includegraphics[width=1.9cm]{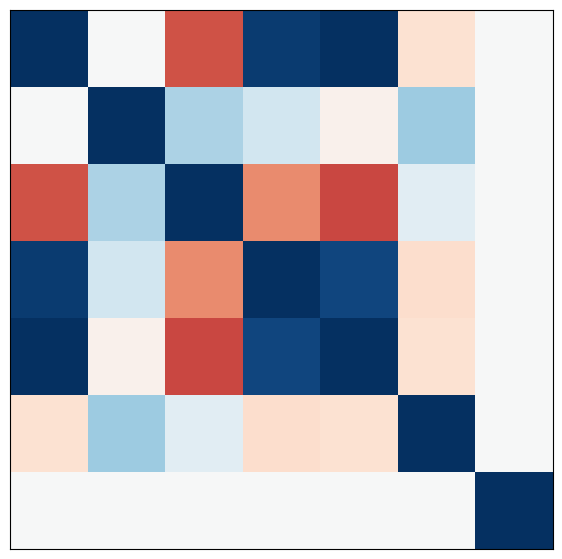}
        \captionsetup{labelformat=empty}
        \vspace{-0.2cm}
        \captionof{subfigure}{2006 BC10}\label{fig:2006_BC10_corr}
    \end{minipage}
    \begin{minipage}[c]{0.19\textwidth}
        \centering
        \includegraphics[width=1.9cm]{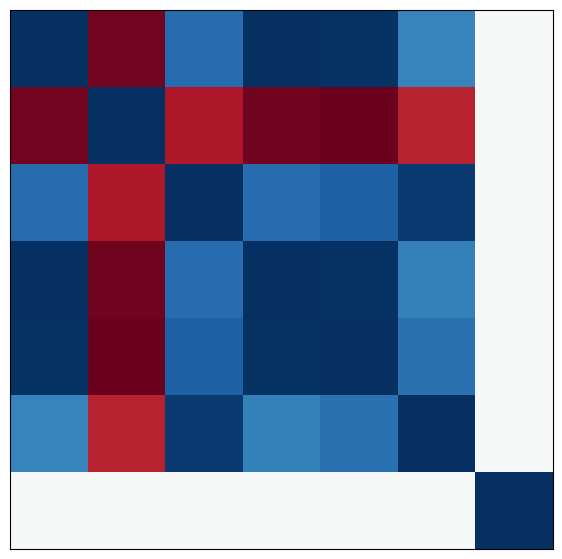}
        \captionsetup{labelformat=empty}
        \vspace{-0.2cm}
        \captionof{subfigure}{2024 QP2}\label{fig:2024_QP2_corr}
    \end{minipage}
    \begin{minipage}[c]{0.19\textwidth}
        \centering
        \includegraphics[width=1.9cm]{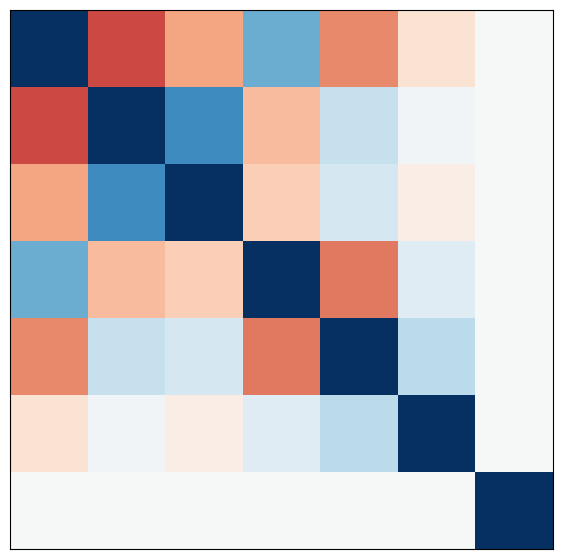}
        \captionsetup{labelformat=empty}
        \vspace{-0.2cm}
        \captionof{subfigure}{2015 DU180}\label{fig:2015_DU180_corr}
    \end{minipage}
    \begin{minipage}[c]{0.19\textwidth}
        \centering
        \includegraphics[width=1.9cm]{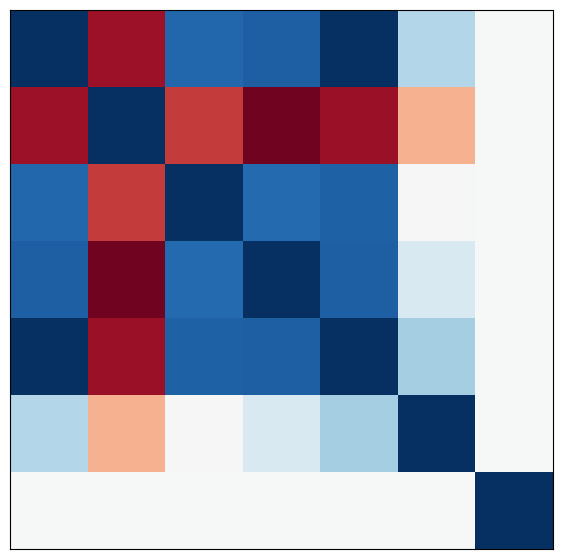}
        \captionsetup{labelformat=empty}
        \vspace{-0.2cm}
        \captionof{subfigure}{2014 WK7}\label{fig:2014_WK7_corr}
    \end{minipage}
    \centering
    \vspace{0.15cm}
    \begin{minipage}[c]{0.4\textwidth}
        \centering
        \includegraphics[width=7cm]{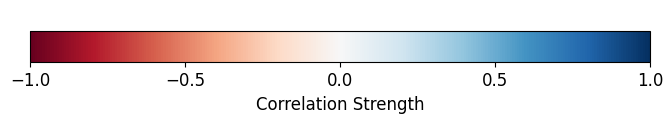}
        \captionsetup{labelformat=empty}
    \end{minipage}
    \caption{Correlation matrices for the sample of 20 observed asteroids during a close approach with one of LISA spacecraft computed through the Fisher information.}
    \label{fig:correlation_matrices_all}
\end{minipage}
\end{figure}

The correlation matrices for the sample of asteroids are displayed in Fig. \ref{fig:correlation_matrices_all}. While the correlation matrices emphasize the uniqueness of each close approach, they also reveal recurring patterns among asteroids. In particular, it can be shown that the observed patterns are linked not just to direct orbital elements, but also to analogous relative velocities and directions during the close approach. This supports the intuition that the structure of the correlation matrices is primarily governed by the flyby geometry, whereas the flyby distance (MOID) predominantly determines the S/N. In addition, the significance of these generalized patterns lies in their potential applicability to nonideal detection scenarios, in which only a limited number of orbital parameters can be determined. In these cases, the characteristic patterns of the correlation matrices could serve as informative priors to guide the later estimation or retrieval of the remaining orbital parameters.

To summarize, although a LISA measurement cannot match the precision of current astrometric methods, it can provide complementary capabilities, in particular, for a mass determination. This would substantially enhance the NEA catalog.


\section{Conclusion}\label{section:conclusion}

This is the first study to systematically assess the signals caused by asteroids encountering the LISA S/C as they appear in the LISA data stream after the removal of laser frequency noise. There are two key conclusions form this study. 

First, the LISA mission can be used to detect and characterize asteroids, provided that they approach one of the spacecraft closely enough. This in itself is an interesting finding, although asteroids have been considered a source of potential noise for LISA, as the mission is designed to detect gravitational waves that affect the distance between the spacecraft. However, the very low residual acceleration noise of the test masses also allows us to assess the effects of static gravity.   

Second, the accuracy with which the mass of such asteroids can be determined in a single encounter is unparalleled by any other current observational methods, except dedicated probes that orbit the asteroid for an extended time. 
As the mass (and subsequently, the composition) of many asteroid is only loosely constrained by first principles and observation, such measurements might be very useful for the assessment of the hazard potential of NEAs and their potential origin.

Although our analysis was exclusively based on the currently observed NEA population, it can be extended to objects that remain undiscovered. In particular, we showed that NEAs with radii of $\sim 500$ \unit{\si{\meter}}  could produce signals that are detectable by LISA if they have certain minimum mass or MOID. The currently cataloged population is estimated to be complete to approximately 80 \% for radii $\geq 500$ \unit{\si{\meter}}  \citep{Grav2023}. This implies that a non-negligible fraction of potentially detectable NEAs is yet to be identified. Consequently, LISA observations may reveal and enable the characterization of previously unknown NEAs within this size range.

We emphasize that for typical close encounters, LISA alone is unlikely to yield orbital solutions or state vectors with the precision required for a comprehensive dynamical characterization, except in particularly favorable high S/N cases such as asteroid 4179 Toutatis. Nevertheless, the localization accuracies achievable in our framework translate into a preliminary orbit determination and a search region in the sky. These error ellipses can, in some cases, substantially reduce this search area for targeted follow-up observations, either from ground-based facilities or from complementary space-borne missions.

Finally, the expected detection rate over the nominal mission lifetime remains modest and does not compromise the primary scientific objectives of LISA. Currently, our most optimistic estimate\footnote{Based on the LISA sensitivity limit below 0.1 \unit{\si{\milli\hertz}}, we established a maximum detection distance of $10^5$ \unit{\si{\kilo\meter}}. With an S/N threshold of 5, the smallest detectable asteroid diameter at this distance is $\sim140$ \unit{\si{\meter}}. For the \citep{Farnocchia_2021} size distribution and accounting for all three spacecraft, the predicted encounter rate is $\sim0.27$ \unit{\year}$^{-1}$.} suggests that about three detections may arise from the currently known population for the three spacecraft, with an additional currently unconstrained contribution from undiscovered objects during a period of 10 years (corresponding to the LISA lifetime). Even at this level, any detection would represent a meaningful addition to the NEA inventory and would demonstrate the contribution of LISA to small-body studies, especially in complement with measurements of poorly characterized physical properties such as the mass.

We only addressed the question how well the parameters for asteroids can be measured in a close encounter with any LISA S/C. The questions how these results, which assumed an optimal measurement environment, relate to the characterization of  the currently observed (and unobserved) population of near-Earth asteroids, the prospects for LISA of finding currently unknown NEAs, or whether their parameters can be extracted from the detected signal are subject for future work. We also still need to determine whether we can learn something about the shape or the density profile of asteroids through their encounter with LISA. This is deferred to a future study.


\begin{acknowledgements} 
The authors would like to thank D.~Koschny for useful discussions regarding the importance of the topic and an introduction into the physics of Near Earth Objects. 

This work made use of the software packages: 
\texttt{pyTDI} \citep{pyTDI}, \texttt{LISA Instrument} \citep{lisainstrument}, \texttt{LISA Constants} \citep{lisaconstants}, \texttt{LISA Glitch} \citep{lisaglitch},
\texttt{Jupyter} \citep{2007CSE.....9c..21P,kluyver2016jupyter}, \texttt{matplotlib} \citep{Hunter:2007}, \texttt{numpy} \citep{numpy}, \texttt{pandas} \citep{mckinney-proc-scipy-2010,pandas_13819579}, \texttt{python} \citep{python}, \texttt{scipy} \citep{2020SciPy-NMeth,scipy_15366870} and \texttt{h5py} \citep{collette_python_hdf5_2014,h5py_7560547}.

Some of the software citation information was aggregated using \texttt{\href{https://www.tomwagg.com/software-citation-station/}{The Software Citation Station}} \citep{software-citation-station-paper,software-citation-station-zenodo}.
\end{acknowledgements}


\onecolumn

\clearpage

\renewcommand{\arraystretch}{1.05}
\begin{table}[htp!]
{
\small
\centering
\caption{S/N and standard deviations of state vector and mass for a sample of 20 observed asteroids for a close approach at the MOID with one of LISA's spacecraft}
{
\begin{tabular}{@{}rcccccccccc@{}}
\toprule
  \multicolumn{1}{c}{Asteroid} & S/N & $\sigma_x$ [\unit{\si{\astronomicalunit}}] & $\sigma_y$ [\unit{\si{\astronomicalunit}}] & $\sigma_z$ [\unit{\si{\astronomicalunit}}] & $\sigma_{v_x}$ [\unit{\si{\kilo\meter\per\second}}] & $\sigma_{v_y}$ [\unit{\si{\kilo\meter\per\second}}] & $\sigma_{v_z}$ [\unit{\si{\kilo\meter\per\second}}] & $\sigma_m^{*}$ [\%] & $\sigma_t$ [\unit{\si{\second}}]\\ 
\cmidrule(r){1-1}\cmidrule(lr){2-2}\cmidrule(lr){3-3}\cmidrule(lr){4-4}\cmidrule(lr){5-5}\cmidrule(lr){6-6}\cmidrule(lr){7-7}\cmidrule(lr){8-8}\cmidrule(lr){9-9}\cmidrule(r){10-10}

 4179 Toutatis & 21970 & $7.65\times10^{-5}$  & $6.59\times10^{-5}$ & $1.54\times10^{-5}$ & $3.13\times10^{-3}$ & $2.56\times10^{-3}$ & $1.11\times10^{-3}$ & $4.55\times10^{-3}$ & 3.77 \\ 

 69230 Hermes & 2928 & $4.23\times10^{-4}$ & $1.08\times10^{-4}$ & $5.81\times10^{-5}$ & $2.56\times10^{-2}$ & $1.33\times10^{-2}$ & $2.79\times10^{-3}$ & $3.41\times10^{-2}$ & 5.56  \\

 2002 PZ39 & 2632 & $4.29\times10^{-4}$ & $7.67\times10^{-4}$ & $7.14\times10^{-5}$ & $2.68\times10^{-2}$ & $4.96\times10^{-2}$ & $1.04\times10^{-3}$ & $3.79\times10^{-2}$ & 4.88  \\
 
 2013 FA8 & 1905  & $1.84\times10^{-3}$ & $6.29\times10^{-4}$ & $2.30\times10^{-4}$ & $2.87\times10^{-2}$ & $0.14$ & $3.25\times10^{-3}$ & $5.31\times10^{-2}$ & 1.28 \\ 

 1997 BR & 1381  & $4.55\times10^{-3}$ & $6.16\times10^{-4}$ & $1.71\times10^{-3}$ & $6.51\times10^{-2}$ & $0.14$ & $1.58\times10^{-2}$ &  $7.25\times10^{-2}$ & 17.02 \\

 2007 UY1 & 1071  & $1.44\times10^{-3}$ & $8.49\times10^{-4}$ & $5.16\times10^{-5}$ & $1.95\times10^{-2}$ & $7.39\times10^{-2}$ & $6.77\times10^{-4}$ &  $9.37\times10^{-2}$ & 2.88 \\

 2003 QQ47 & 956 & $2.67\times10^{-3}$ & $1.40\times10^{-3}$ & $8.44\times10^{-4}$ & $3.92\times10^{-2}$ & $4.57\times10^{-2}$ & $8.75\times10^{-2}$ &  $0.10$ & 30.55 \\

 1999 MN & 886  & $5.86\times10^{-4}$ & $8.88\times10^{-4}$ & $1.54\times10^{-5}$ & 0.18 & 0.21 & $1.76\times10^{-2}$ &  $0.11$ & 9.16  \\ 

 2003 YK118 & 743 & $9.70\times10^{-4}$  & $1.54\times10^{-3}$ & $5.63\times10^{-4}$ & $8.11\times10^{-2}$  & $2.16\times10^{-2}$ & $1.94\times10^{-2}$ &  $0.14$ & 80.47 \\

 2011 BO24 & 442 & $8.35\times10^{-3}$  & $1.34\times10^{-4}$ & $2.60\times10^{-3}$ & $3.75\times10^{-2}$  & $0.25$ & $5.69\times10^{-2}$ & $0.23$ & 94.63\\ 

 2003 AC23 & 349 & $3.30\times10^{-3}$ & $7.58\times10^{-3}$ & $1.73\times10^{-4}$ & $0.35$ & $4.58\times10^{-2}$ & $1.07\times10^{-2}$ &  $0.29$ & 21.35 \\ 

 2020 FR1 & 229 & $1.95\times10^{-2}$ & $4.95\times10^{-3}$ & $4.00\times10^{-3}$ & 0.85 & 1.63 & 0.20 & $0.44$ & 38.64 \\ 

 2018 DA1 & 209 & $4.71\times10^{-3}$ & $4.16\times10^{-3}$ & $4.13\times10^{-3}$ & 0.57 & 0.87 & 0.46 & $0.48$ & 125.47 \\ 

 2016 CU193 & 155 & $2.15\times10^{-2}$ & $2.58\times10^{-3}$ & $1.19\times10^{-2}$ & 0.49 & 0.68 & 0.42 & $0.65$ & 83.84 \\

 2014 UC192 & 118 & $7.51\times10^{-3}$  & $1.27\times10^{-2}$ & $2.05\times10^{-3}$ & 0.56 & 0.18 & $7.48\times10^{-2}$ & 0.85 & 29.80 \\

 2005 TU50 & 96 & $2.59\times10^{-2}$ & $1.39\times10^{-2}$ & $6.58\times10^{-3}$ & 1.10 & 2.03 & 0.40 & $1.04$ & 128.55 \\

 2006 BC10 & 32 & $8.00\times10^{-2}$ & $1.79\times10^{-2}$ & $9.86\times10^{-3}$ & 2.17 & 4.82 & $0.37$ &  $3.16$ & 2680.03 \\

 2024 QP2 & 24 & $4.59\times10^{-2}$ & $3.55\times10^{-2}$ & $7.00\times10^{-2}$ & 1.65 & 1.65 & 0.59 & $4.22$ & 1598.61 \\

 2015 DU180 & 11 & 0.14 & 0.11 & $5.30\times10^{-2}$ & 72.61 & 128.17 & 31.16 & 8.71 & 5523.18 \\

 2014 WK7 & 5 & 0.27 & 0.20 & 0.10 & 6.73 & 9.90 & 3.51 & 18.50 & 4668.18 \\
 
\bottomrule
\end{tabular}}
\tablefoot{The asteroids are organized by descending order of S/N. Note that the mass standard deviation, $\sigma_m^*$, corresponds to the normalized version $\sigma_m^* = \sigma_m/m$ by the mass $m$ of the asteroid.}
\label{table:standard_deviations_all}
}
\end{table}

\twocolumn


\bibliographystyle{aa}
\bibliography{bibliography}



\clearpage
\onecolumn

\begin{appendix}

\section{Properties of investigated near-Earth asteroids}
\renewcommand{\thetable}{A.\arabic{table}}
\setcounter{table}{0}

\vspace{0.5cm}
In this appendix, we provide the orbital elements and computed physical properties of the sample of NEAs from the observed population used throughout the simulations.
\vspace{0.5cm}

\renewcommand{\arraystretch}{1.2}
\begin{table}[htp!]
{
\small
\centering
\caption{Orbital elements and physical properties for the sample of 20 observed asteroids collected at epoch 61200 mjd.}
\begin{tabular}{@{}rccccccccccccc@{}}
\toprule
  \multicolumn{1}{c}{Asteroid} & $a$ [\unit{\si{\astronomicalunit}}] & $e$ & $i$ [$^{\circ}$] & $\Omega$ [$^{\circ}$] & $\omega$ [$^{\circ}$] & H [mag] & albedo & $d$ [km] & $m$ [kg] & MOID [\unit{\si{\astronomicalunit}}] & $M_\text{MOID}$ [rad] \\ 
\cmidrule(r){1-1}\cmidrule(lr){2-2}\cmidrule(lr){3-3}\cmidrule(lr){4-4}\cmidrule(lr){5-5}\cmidrule(lr){6-6}\cmidrule(lr){7-7}\cmidrule(lr){8-8}\cmidrule(lr){9-9}\cmidrule(lr){10-10}\cmidrule(lr){11-11}\cmidrule(l){12-12}

 4179 Toutatis & 2.54 & 0.62 & 0.45 & 125.37 & 277.86 & 15.30 & 0.254 & 2.64 & $2.12\times10^{13}$ & $4.10\times10^{-4}$ & $(6.19,\,\, 2.74)$ \\ 

 69230 Hermes & 1.66 & 0.62 & 6.07 & 34.03 & 92.93 & 17.57 & 0.265 & 0.82 & $6.35\times10^{11}$ & $1.45\times10^{-4}$ & $(0.40,\,\, 2.07)^{*}$ \\ 

 2002 PZ39 & 1.47 & 0.55 & 1.67 & 328.72 & 260.19 & 19.10 & - & - & $1.79\times10^{11}$ & $9.89\times10^{-5}$ & $(0.51,\,\, 3.83)^{*}$ \\

 2013 FA8 & 1.93 & 0.67 & 3.12 & 165.42 & 281.06 & 20.93 & - & - & $1.43\times10^{10}$ & $2.28\times10^{-5}$ & $(0.30,\,\, 5.50)$ \\ 

 1997 BR & 1.34 & 0.31 & 17.25 & 116.63 & 133.86 & 17.80 & 0.51 & 0.56 & $2.02\times10^{11}$ & $1.31\times10^{-4}$ & $(0.42,\,\, 5.56)^{**}$ \\

 2007 UY1 & 0.95 & 0.18 & 1.02 & 337.45 & 274.12 & 23.08 & - & - & $7.28\times10^{8}$ & $7.64\times10^{-6}$ & $(4.60,\,\, 0.68)^{*}$ \\ 

 2003 QQ47 & 1.09 & 0.19 & 62.10 & 0.98 & 104.97 & 17.37 & - & - & $1.96\times10^{12}$ & $4.38\times10^{-4}$ & $(0.96,\,\, 1.46)^{*}$ \\ 

 1999 MN & 0.67 & 0.67 & 2.03 & 81.38 & 9.21 & 20.57 & - & - & $2.35\times10^{10}$ & $6.29\times10^{-5}$ & $(1.90,\,\, 0.57)$ \\

 2003 YK118 & 1.70 & 0.49 & 7.83 & 326.73 & 233.28 & 18.81 & 0.07 & 0.90 & $8.40\times10^{11}$ & $3.44\times10^{-4}$ & $(5.97,\,\, 5.05)$ \\
 
 2011 BO24 & 1.38 & 0.28 & 20.37 & 91.46 & 178.67 & 18.78 & - & - & $2.78\times10^{11}$ & $2.42\times10^{-4}$ & $(0.01,\,\, 5.14)^{**}$ \\ 

 2003 AC23 & 2.17 & 0.58 & 2.06 & 43.73 & 126.41 & 21.90 & - & - &  $3.72\times10^{9}$ & $3.97\times10^{-5}$ &  $(0.16,\,\, 6.17)$ \\ 

 2020 FR1 & 1.41 & 0.61 & 13.11 & 26.37 & 255.89 & 21.20 & - & - & $9.86\times10^{9}$ & $8.58\times10^{-5}$ & $(0.54,\,\, 2.92)$ \\ 

 2018 DA1 & 1.21 & 0.73 & 25.49 & 139.90 & 233.94 & 18.85 & - & - & $2.52\times10^{11}$ & $3.37\times10^{-4}$ & $(0.62,\,\, 4.95)$ \\ 

 2016 CU193 & 1.47 & 0.36 & 49.89 & 322.37 & 323.12 & 19.94 & - & - & $5.59\times10^{10}$ & $1.73\times10^{-4}$ & $(0.29,\,\, 1.83)$ \\

 2014 UC192 & 1.41 & 0.36 & 11.66 & 34.92 & 304.61 & 23.72 & - & - & $3.03\times10^{8}$ & $1.78\times10^{-5}$ & $(0.44,\,\, 3.06)$ \\

 2005 TU50 & 1.43 & 0.60 & 12.42 & 22.18 & 259.71 & 21.45 & - & - & $6.92\times10^{9}$ & $1.14\times10^{-4}$ & $(0.53,\,\, 2.84)$\\

 2006 BC10 & 2.02 & 0.66 & 0.91 & 20.25 & 235.24 & 19.50 & - & - & $1.03\times10^{11}$ & $7.24\times10^{-4}$ & $(6.01,\,\, 5.62)$\\

 2024 QP2 & 2.54 & 0.62 & 1.64 & 205.32 & 202.15 & 21.57 & - & - & $5.83\times10^{9}$ & $1.78\times10^{-4}$ & $(6.20,\,\, 4.98)^{*}$\\

 2015 DU180 & 1.93 & 0.92 & 4.86 & 51.74 & 315.41 & 21.03 & 0.28 & 0.39 & $6.83\times10^{10}$ & $1.08\times10^{-3}$ & $(6.05,\,\, 0.17)$ \\
 
 2014 WK7 & 1.56 & 0.36 & 23.91 & 233.74 & 180.93 & 22.17 & - & - & $2.59\times10^{9}$ & $1.68\times10^{-4}$ & $(6.27,\,\, 1.32)^{**}$\\
 
\bottomrule
\end{tabular}
\tablefoot{The mass was computed using Eq. \ref{equation:mass} either using the values of observed properties or assuming the intermediate values. $M_\text{MOID}$ indicates the mean anomalies needed for the asteroid and spacecraft 2 ($^{*}$ for spacecraft 1, $^{**}$ for spacecraft 3) to be in a close approach configuration at the MOID, respectively.}
\label{table:orbital_elements_NEA}
}
\end{table}

\end{appendix}

\end{document}